\documentclass[pra,showpacs,onecolumn]{revtex4}
\usepackage{graphicx}
\usepackage{dcolumn}
\usepackage{bm}
\usepackage{amssymb}
\usepackage{amsmath}
\usepackage{epsfig}

\input{tcilatex}

\begin{document}

\title{Permutation asymmetry inducing entanglement between degrees of
freedom in multiphoton states}
\author{F. W. Sun,$^{1,2,} \footnote{%
E-mail: fs2293@columbia.edu}$ B. H. Liu,$^{2}$ C. W. Wong,$^{1}$ and G. C.
Guo$^{2}$ }
\affiliation{$^{1}$Optical Nanostructures Laboratory, Columbia University, New York, NY
10027. \\
$^{2}$Key Laboratory of Quantum Information, University of Science and
Technology of China, CAS, Hefei, 230026, the People's Republic of China. }
\date{\today }

\begin{abstract}
We describe and examine entanglement between different degrees of freedom in
multiphoton states based on the permutation properties. From the state
description, the entanglement comes from the permutation asymmetry.
According to the different permutation properties, the multiphoton states
can be divided into several parts. It will help to deal with the multiphoton
interference, which can be used as the measurement of the entanglement.
\end{abstract}

\maketitle

\section{Introduction}

Photons interference has been widely applied in different protocols of
quantum communication \cite{BB84,Gisin}, quantum computation \cite{Kok}, and
quantum metrology \cite{Fonseca,D'Angelo,Walther,mitchell}. Many of those
protocols utilize one degree of freedom (DOF), such as polarization, time
(energy), momentum (path), etc., and neglect the relationship with others
due to either filtering or the absence of correlations. Recently, however,
more and more cases are discussed with multiple DOFs in the system, where a
multidimensional correlated system can be remarkably formed \cite{Moreva}.
In some cases, there is no entanglement between different DOFs, such as the
hyperentanglement state \cite{Barreiro,Cinelli,Barbieri}. In others, there
may be entanglement \cite{Moreva}. In those states, entanglement will bring
decoherence for one DOF when the other DOFs are discarded. It has been
well-described and observed for the two-photon correlated system \cite%
{Adamson}.

Moreover, there are many cases focusing on the multiphoton system. The
relationships between different DOFs and different photons are significantly
more complicated than the two-photon cases. This is especially true for
photons generated from parametric down conversion (PDC), where it is
difficult to describe the state of more than one photon in the same mode of
spatial DOF. There has been several experiments discussing the decoherence
in the polarization DOF in the four-photon state from PDC \cite%
{Tsujino,Eisenberg}. The interpretation based on photons distinguishability
has been well proposed \cite{Ou99,Ou05}. Recently, a description based on
photon permutation symmetry was proposed \cite{Sun07}, where it described
that entanglement between two DOFs decreased state purity and interference
visibility. However, in that approach, it is not convenient to tell if there
is entanglement between different DOFs by fully decomposing the state.
Moreover, the description is appreciably more involved when there are
multiphoton states in the same mode of spatial DOF.

In this paper, we will develop and introduce a convenient approach to
determine the entanglement between different DOFs in a multiphoton state.
The method is based on the permutation symmetry of different DOFs in the
state, extending from permutation symmetry of different photons. Generally,
the photons in one DOF are distinguishable because they can be recognized by
the information of the other DOF. There are correlations between the two
DOFs, which we call entanglement. On the other hand, the indistinguishable
photons in one DOF should have a permutation symmetric form. Therefore, the
entanglement is induced by the permutation asymmetry and can be read
directly from the state which is described by photon creation operators. If
there is no entanglement between different DOFs, the state can be described
in a product form. Thus, a single DOF in different states will show the same
interference behavior under the same operation if their descriptions are the
same. For example, both the Greenberger-Horne-Zeilinger state (GHZ) \cite%
{Greenberger} and the maximally entangled number state (NOON) \cite%
{Bollinger,Sun06} can be applied in the demonstration of the multiphoton de
Broglie wavelength and the high resolution quantum phase measurement to
approach the Heisenberg limit \cite{Bollinger,Sun06PRA}. They will be
described in Sec. II. When there is entanglement between different DOFs, the
single DOF will not show perfect interference. In Sec. III, a four-photon
state interference in the same spatial mode, which can be used as the
measurement of the entanglement, will be described in detail. The
four-photon state will be divided in several parts according to their
permutation symmetries. It is much more convenient than other methods \cite%
{Sun06PRA,Sun07}. The last section is the conclusion.

\section{Multiphoton multi-DOF entanglement}

To clearly describe the interference, the multiphoton state is written in a
permutation symmetric form \cite{Sun07}

\begin{equation}
\dprod\limits_{i=1}^{N}a_{i}^{\dag }\left\vert \text{vac}\right\rangle =%
\frac{1}{\sqrt{N!}}\sum_{P}P(\left\vert a_{1}\right\rangle \left\vert
a_{2}\right\rangle \cdot \cdot \cdot \left\vert a_{N}\right\rangle )\text{,}
\label{nphoton}
\end{equation}%
where $P$ is the permutation operator that changes the positions of
arbitrary two states. There are $N!$ terms for the $N$ photons. For example,
a two-photon state can be described as $a_{1}^{\dag }a_{2}^{\dag }\left\vert
\text{vac}\right\rangle =\sum_{P}P(\left\vert a_{1}\right\rangle \left\vert
a_{2}\right\rangle )/\sqrt{2!}=(\left\vert a_{1}\right\rangle \left\vert
a_{2}\right\rangle +\left\vert a_{2}\right\rangle \left\vert
a_{1}\right\rangle )/\sqrt{2}$. For the identity case $\left\vert
a_{1}\right\rangle =\left\vert a_{2}\right\rangle =\left\vert a\right\rangle
$, $a^{\dag 2}\left\vert \text{vac}\right\rangle =\sqrt{2}\left\vert
a\right\rangle \left\vert a\right\rangle $. If there is more than one
uncoupled DOF in the system, the single-photon state is described as a
product of single DOF,
\begin{equation}
a^{\dag }(\alpha ,\beta ,...,\gamma )\left\vert \text{vac}\right\rangle
=\left\vert \alpha ,\beta ,...,\gamma \right\rangle =\left\vert \alpha
\right\rangle \left\vert \beta \right\rangle \cdot \cdot \cdot \left\vert
\gamma \right\rangle \text{,}  \label{nDOF}
\end{equation}%
where $\alpha $, $\beta $, ..., $\gamma $ represent the different DOFs.

Based on Eqs. (\ref{nphoton}) and (\ref{nDOF}), we can discuss the
entanglement between different DOFs in a multiphoton state. The examples of
two-photon and four-photon states with two DOFs have been presented in Ref.
\cite{Sun07}. Generally, the $N$-photon state containing two DOFs can be
written as \cite{note}
\begin{equation}
\left\vert \Psi _{N}\right\rangle =\sum_{\alpha _{1},\beta _{1},...,\alpha
_{N},\beta _{N}}f(\alpha _{1},\beta _{1},...,\alpha _{N},\beta
_{N})\prod\limits_{k=1}^{N}a^{\dagger }(\alpha _{k},\beta _{k})\left\vert
\text{vac}\right\rangle \text{,}  \label{N-photon}
\end{equation}%
where $\alpha $ and $\beta $ are the two DOFs.

In general, if there is no entanglement between the two DOFs, each DOF will
have a permutation symmetric form. This is a result of permutation symmetry
of bosonic particles. Thus, with this underlying principle, we can tell the
entanglement based on the permutation symmetry of the state description.

Under the permutation of any photon's total wave function, the state
described in Eq. (\ref{N-photon})\ is invariant \cite{Loudon}, such that
\begin{equation}
f(...,\alpha _{i},\beta _{i},...,\alpha _{j},\beta _{j},...)\ =f(...,\alpha
_{j},\beta _{j},...,\alpha _{i},\beta _{i},...)\text{.}\   \label{C0}
\end{equation}%
For a fixed set of $\alpha $, if there is any permutation of the other DOF ($%
\beta $) satisfying
\begin{equation}
f(...,\alpha _{i},\beta _{i},...,\alpha _{j},\beta _{j},...)\ =f(...,\alpha
_{i},\beta _{j},...,\alpha _{j},\beta _{i},...)\text{,}\   \label{C1}
\end{equation}%
this part of the state can then be written as a product of permutation
symmetric states
\begin{eqnarray}
&&\sum_{\beta _{1},...,\beta _{N}}f(\alpha _{1},\beta _{1},...,\alpha
_{N},\beta _{N})\prod\limits_{k=1}^{N}a^{\dagger }(\alpha _{k},\beta
_{k})\left\vert vac\right\rangle   \notag \\
&\rightarrow &\sum_{P}(\left\vert \alpha _{1}\right\rangle \left\vert \alpha
_{2}\right\rangle ,...,\left\vert \alpha _{N}\right\rangle ){\Big (}%
\sum_{\beta _{1},...,\beta _{N}}f(\alpha _{1},\beta _{1},...,\alpha
_{N},\beta _{N})  \notag \\
&&\times \sum_{P}P(\left\vert \beta _{1}\right\rangle \left\vert
\beta _{2}\right\rangle ,...,\left\vert \beta _{N}\right\rangle
){\Big )}\text{.} \label{P1}
\end{eqnarray}%
If $\beta $ DOF keeps the same description (described in the large
parentheses in the above expression) for all of the permutation states of $%
\alpha $ DOF, the whole state can be written in a product form and there is
no entanglement between the two DOFs. Otherwise, there is entanglement.
Equation (\ref{C0}) describes the permutation symmetry of whole wave
function for bosonic particles, while Eq. (\ref{C1}) describes the
permutation symmetry of single DOF. It is a necessary condition for that
there is no entanglement between different DOFs. The examples of two-photon
states are discussed in detail in Ref. \cite{Sun07}. We note that the Bell
singlet state is a special case in which both DOFs are in permutation
antisymmetric form.

When there is no entanglement between different DOFs, a collective (on all
qubits) operation on one DOF will have no effect on the other DOF and the
photons in the DOF\ will show perfect interference. Moreover, if the
description of one DOF is the same, the behavior under the same operation
will be the same too.

For the multiphoton polarized state, if all photons are in the same mode of
spatial DOF, there is no entanglement between the polarization DOF and the
spatial DOF. The whole photon state can be written in a product form. For
example, the NOON state is described as
\begin{eqnarray}
\left\vert \text{NOON}\right\rangle  &=&(a_{H,S}^{\dag N}+a_{V,S}^{\dag
N})\left\vert \text{vac}\right\rangle /\sqrt{2N!}  \notag \\
&=&(\left\vert H\right\rangle ^{\otimes N}+\left\vert V\right\rangle
^{\otimes N})\otimes \left\vert S\right\rangle ^{\otimes N}/\sqrt{2}\text{,}
\label{NOON}
\end{eqnarray}%
where $S$ is the spatial mode and $H$ and $V$ are horizontal and vertical
polarizations, respectively. In addition, when $N$ photons are different
spatial modes, there also exists a product state that has no entanglement
between two DOFs, such as the GHZ state,
\begin{eqnarray}
\left\vert \text{GHZ}\right\rangle
&=&(\dprod\limits_{i=1}^{N}a_{H,S_{i}}^{\dag
}+\dprod\limits_{i=1}^{N}a_{V,S_{i}}^{\dag })\left\vert \text{vac}%
\right\rangle /\sqrt{2}  \notag \\
&=&(\left\vert H\right\rangle ^{\otimes N}+\left\vert V\right\rangle
^{\otimes N})\otimes \sum_{P}P(\left\vert S_{1}\right\rangle \left\vert
S_{2}\right\rangle \cdot \cdot \cdot \left\vert S_{N}\right\rangle )/\sqrt{%
2N!}\text{,}  \label{GHZ}
\end{eqnarray}%
where $S_{i}$ are for the $i$th spatial modes. As shown in Eqs. (\ref{NOON})
and (\ref{GHZ}), both the polarization DOF and the spatial DOF have the
permutation symmetric form. Moreover, in the NOON state and $N$-photon GHZ
state, the polarization DOF has the same form. If an operation acts
collectively on this DOF, the two states will show the same results. For
example, both of them will show the same application in the quantum phase
measurement.
\begin{figure}[h]
\includegraphics[width=8cm]{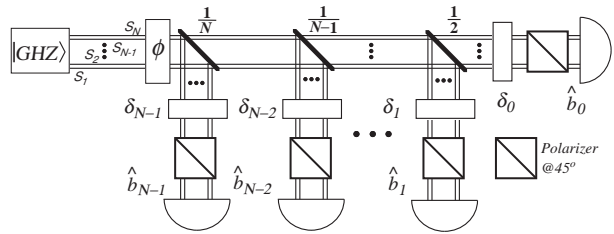}
\caption{Illustration of NOON state projection on a GHZ state. $\protect\phi
$ is the relative phase shift between two polarizations. Number above each
beam splitter denotes the reflectivity. $\protect\delta _{k}=2k\protect\pi /N
$ is the phase delay between $H$ and $V$ polarizations. The polarizers are $%
45^{\circ }$ oriented \protect\cite{Sun06}. The detectors cover all $N$
spatial modes.}
\label{NOON-GHZ}
\end{figure}

As we know, the NOON \cite{Bollinger} state is a popular state for the
quantum phase measurement. In the process, there is a relative phase shift $%
\phi $ between the two polarizations on all $N$ photons which will cause a
whole phase shift $e^{iN\phi }$. The result of NOON state projection \cite%
{Sun06,Sun06PRA,Resch} on the two states will show cosinusoidal oscillation
for the $N$-photon de Broglie wavelength. In the NOON state projection
measurement for the GHZ state, as shown in Fig. \ref{NOON-GHZ}, each
single-photon detector covers all $N$ spatial modes. The $N$-fold
coincidence counts will show the successful projection measurement, which
can be described as

\begin{equation}
M=(\left\vert H\right\rangle ^{\otimes N}-e^{-iN\phi }\left\vert
V\right\rangle ^{\otimes N})(\left\langle H\right\vert ^{\otimes
N}-e^{iN\phi }\left\langle V\right\vert ^{\otimes N})\otimes I_{P}\text{,}
\end{equation}%
where $I_{P}=$ $\sum_{P}P(\left\vert S_{1}\right\rangle \left\vert
S_{2}\right\rangle \cdot \cdot \cdot $ $\left\vert S_{N}\right\rangle
\left\langle S_{1}\right\vert \left\langle S_{2}\right\vert \cdot \cdot
\cdot $ $\left\langle S_{N}\right\vert )/N!$ is the matrix for the spatial
DOF. Here we neglect the total coefficient from the photon loss of each beam
splitter. The measurement result is

\begin{equation}
R=\left\langle \text{GHZ}\right\vert M\left\vert \text{GHZ}\right\rangle
=(1-\cos N\phi )/2\text{.}
\end{equation}%
Thus, the GHZ state can show the oscillation of de Broglie wavelength
behavior. It can also be applied to the high resolution quantum phase
measurement to approach the Heisenberg limit.

\section{Four-photon interference visibility with entangled DOFs}

When there is entanglement between two DOFs, there will be
distinguishability in one DOF, and will not be perfect interference in this
DOF. The character of permutation symmetry can help to describe the
distinguishability in the photon interference, even to calculate the
interference visibility. Here we will utilize the different permutation
properties to divide the state into several parts. The visibility
calculation is much simplified with this method. As an example, we will
discuss the four-photon interference in a single mode of spatial DOF.

As described in Refs. \cite{Sun06,Sun06PRA}, the two-photon state from the
two-cascaded type-I BBOs is expressed as
\begin{eqnarray*}
\left\vert \Psi _{2}\right\rangle  &=&\frac{1}{\sqrt{2}}\dsum\limits_{\alpha
}\varphi (\alpha )[a^{\dag 2}(H,\alpha )+a^{\dag 2}(V,\alpha )]\left\vert
\text{vac}\right\rangle  \\
&=&\frac{1}{\sqrt{2}}(\left\vert HH\right\rangle +\left\vert
VV\right\rangle ){\Big (}\dsum\limits_{\alpha }\varphi (\alpha
)\left\vert \alpha \alpha \right\rangle {\Big )}\text{,}
\end{eqnarray*}%
where $\alpha $ is for another DOF, such as frequency DOF. For simplicity,
we assume $\varphi (\alpha )$ is real and $\sum_{\alpha }\varphi ^{2}(\alpha
)=1$.

Correspondingly, the four-photon state is

\begin{equation}
\left\vert \Psi _{4}\right\rangle =\frac{1}{2}{\Big
(}\dsum\limits_{\alpha
}\varphi (\alpha )[a^{\dag 2}(H,\alpha )+a^{\dag 2}(V,\alpha )]{\Big )}%
^{2}\left\vert \text{vac}\right\rangle \text{.}
\end{equation}%
However, there is a permutation asymmetric part in the above four-photon
state, which induces the entanglement between the two DOFs. The state can be
rewritten into two parts,
\begin{equation}
\left\vert \Psi _{4}\right\rangle =\text{$\frac{1}{2}$}(\left\vert \Psi
_{4}\right\rangle _{A}+\left\vert \Psi _{4}\right\rangle _{B})\text{,}
\end{equation}%
where%
\begin{eqnarray}
\left\vert \Psi _{4}\right\rangle _{A} &=&\dsum\limits_{\alpha }\varphi
(\alpha )[a^{\dag 2}(H,\alpha )a^{\dag 2}(H,\alpha )+a^{\dag 2}(V,\alpha
)a^{\dag 2}(V,\alpha )  \notag \\
&&+a^{\dag 2}(H,\alpha )a^{\dag 2}(V,\alpha )+a^{\dag 2}(H,\alpha )a^{\dag
2}(V,\alpha )]\left\vert \text{vac}\right\rangle   \notag \\
&&+\dsum\limits_{\alpha \neq \beta }\varphi (\alpha )\varphi (\beta
)[a^{\dag 2}(H,\alpha )a^{\dag 2}(H,\beta )+a^{\dag 2}(V,\alpha )a^{\dag
2}(V,\beta )]\left\vert \text{vac}\right\rangle   \notag \\
&=&(\left\vert HHHH\right\rangle +\left\vert VVVV\right\rangle ){\Big (}%
\dsum\limits_{\alpha }\sqrt{24}\varphi ^{2}(\alpha )\left\vert \alpha \alpha
\alpha \alpha \right\rangle   \notag \\
&&+\dsum\limits_{\alpha <\beta }\varphi (\alpha )\varphi (\beta
)\sum_{P}P(\left\vert \alpha \alpha \beta \beta \right\rangle )/\sqrt{6}%
{\Big )}  \notag \\
&&+\sum_{P}P(\left\vert HHVV\right\rangle /\sqrt{6})\dsum\limits_{\alpha
}\varphi ^{2}(\alpha )\left\vert \alpha \alpha \alpha \alpha \right\rangle
\label{SYM}
\end{eqnarray}%
is the permutation symmetric part, and
\begin{eqnarray}
\left\vert \Psi _{4}\right\rangle _{B} &=&\dsum\limits_{\alpha \neq \beta
}\varphi (\alpha )\varphi (\beta )[a^{\dag 2}(H,\alpha )a^{\dag 2}(V,\beta
)+a^{\dag 2}(H,\beta )a^{\dag 2}(V,\alpha )]\left\vert \text{vac}%
\right\rangle   \notag \\
&=&2\dsum\limits_{\alpha <\beta }\varphi (\alpha )\varphi (\beta )[a^{\dag
2}(H,\alpha )a^{\dag 2}(V,\beta )+a^{\dag 2}(H,\beta )a^{\dag 2}(V,\alpha
)]\left\vert \text{vac}\right\rangle   \label{ASYM}
\end{eqnarray}%
is the permutation asymmetric part because of the absence of the photon
state $a^{\dag }(H,\alpha )a^{\dag }(V,\alpha )a^{\dag }(H,\beta )a^{\dag
}(V,\beta )\left\vert \text{vac}\right\rangle $. In Eq. (\ref{SYM}), the
permutation symmetry state $\sum_{P}P(\left\vert iijj\right\rangle
)=4(\left\vert iijj\right\rangle +\left\vert ijij\right\rangle +\left\vert
ijji\right\rangle +\left\vert jjii\right\rangle +\left\vert
jiji\right\rangle +\left\vert jiij\right\rangle )$ is from the 24
permutation terms of $iijj$.

\begin{figure}[h]
\includegraphics[width=6cm]{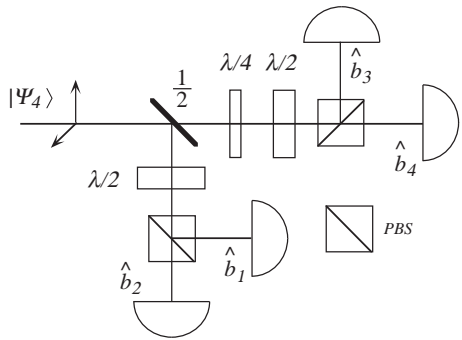}
\caption{NOON state projection measurement for the four-photon state. $%
\protect\lambda /2$ and $\protect\lambda /4$ are half-wave-plate at 22.5$%
^{\circ }$ and quarter-wave-plate at 0$^{\circ }.$}
\label{NOON2}
\end{figure}

The mode in polarization DOF of each photon in $\left\vert \Psi
_{4}\right\rangle _{A}$ is indistinguishable, while it is distinguishable in
$\left\vert \Psi _{4}\right\rangle _{B}$. Thus, we can calculate the results
of the two parts separately. For $\left\vert \Psi _{4}\right\rangle _{A}$,
the polarized NOON state projection measurement \cite{Sun06}, as shown in
Fig. \ref{NOON2}, has the form
\begin{equation}
M=(\left\vert HHHH\right\rangle -\left\vert VVVV\right\rangle )(\left\langle
HHHH\right\vert -\left\langle VVVV\right\vert )\otimes I_{\alpha ,\beta }%
\text{,}
\end{equation}%
where $I_{\alpha ,\beta }$ is the identity matrix for the other DOF because
there is no projection on this DOF in the measurement. This measurement is
constructed based on Hanbury Brown-Twiss interferometer \cite{HBT} by adding
polarization projection before each detector. It is orthogonal with $%
\sum_{P}P(\left\vert HHVV\right\rangle /\sqrt{6})$ and will give null
output, which is the result of Hong-Ou-Mandel interference for multiphoton
\cite{Sun06}. After a phase shift $\phi $, the measurements will show the
perfect interference result, with oscillation for $N$-photon de Broglie
wavelength for $\left\vert \Psi _{4}\right\rangle _{A}$, which are
\begin{eqnarray}
R_{A} &=&_{A}\left\langle \Psi _{4}\right\vert M\left\vert \Psi
_{4}\right\rangle _{A}  \notag \\
&=&2(1-\cos 4\phi )||\dsum\limits_{\alpha }\sqrt{24}\varphi ^{2}(\alpha
)\left\vert \alpha \alpha \alpha \alpha \right\rangle +\dsum\limits_{\alpha
<\beta }\varphi (\alpha )\varphi (\beta )\sum_{P}P(\left\vert \alpha \alpha
\beta \beta \right\rangle )/\sqrt{6}||^{2}  \notag \\
&=&(1-\cos 4\phi )(16+32K)\text{,}
\end{eqnarray}%
where $K=\dsum_{\alpha }\varphi ^{4}(\alpha )\leq 1$ indicates the
entanglement of this DOF in the two-photon state \cite{Sun07,Sun07PRA}.

The permutation asymmetric part $\left\vert \Psi _{4}\right\rangle _{B}$
cannot be written in the product state. There are six cases for the
four-photon state to be detected by four detectors as shown in Fig. \ref%
{NOON2}. However, the case that two $H$ photons are detected by $\hat{b}_{1}%
\hat{b}_{2}$ and two $V$ photons by $\hat{b}_{3}\hat{b}_{4}$ is canceled by
the case of two $H$ photons detected by $\hat{b}_{3}\hat{b}_{4}$ and two $V$
photons by $\hat{b}_{1}\hat{b}_{2}$, since they have contrary phases.
Therefore the result of NOON state projection measurement is not difficult
to calculate
\begin{equation}
R_{C}=\left\Vert \left\vert \Psi _{4}\right\rangle _{B}\right\Vert ^{2}%
{\Large \times }2/3=32(1-K){\Large \times }2/3\text{.}
\end{equation}%
Thus, the total result is
\begin{eqnarray}
R &=&R_{A}+R_{B}  \notag \\
&=&16(7+2K){\Big (}1-\frac{3(1+2K)}{7+2K}\cos 4\phi {\Big
)}/3\text{.} \label{V4}
\end{eqnarray}%
The interference visibility is $V=3(1+2K)/(7+2K)$. It is the same with the
result of Eq. (75) in Ref. \cite{Sun06} if we set $K=\mathcal{E}/\mathcal{A}$%
. Therefore, the higher $K$, the higher interference visibility and the less
entanglement between the two DOFs. When $K=1$, the visibility is $100\%$. In
this case, there is no asymmetric part $\left\vert \Psi _{4}\right\rangle
_{B}$ in the four-photon state and no entanglement between two DOFs.
Therefore, this interferometric method can be used as the measurement of the
entanglement.

\section{Conclusion}

Based on permutation symmetry, we discussed the entanglement between
different DOFs in a multiphoton state. Permutation asymmetry in the state
description induces the entanglement between different DOFs. If a DOF does
not entangle with other DOFs, the same state in this DOF will show the same
interference behavior when an operation acts on all the photons
collectively. As an example, we described that the GHZ state can also be
used to approach Heisenberg limit in the quantum phase measurement. For the
state which has entanglement between different DOFs, there is no maximal
interference in one DOF. The visibility can be calculated by dividing the
state into different parts according to their permutation properties. This
method allows for the description of interference visibility significantly
more conveniently in multiphoton multi-DOF states. Moreover, the
interference can be used as the measurement of the entanglement.

\begin{acknowledgments}
This work is funded by DARPA, NSF Contract No. ECCS 0747787, and the New
York State Office of Science, Technology and Academic Research. B.H.L. and
G.C.G. are supported by Chinese National Fundamental Research Program, and
the National Natural Science Foundation of China.
\end{acknowledgments}


\begin{thebibliography}{99}
\bibitem{BB84} C. H. Bennett and G. Brassard, \textit{Proceedings of the
IEEE International Conference on Computers, Systems, and Signal Processing}
(IEEE, New York, 1984), pp. 175--179.

\bibitem{Gisin} N. Gisin, G. Ribordy, W.Tittel, and H. Zbinden Rev. Mod.
Phys. \textbf{74}, 145 (2002).

\bibitem{Kok} P. Kok, \textit{et al., }Rev. Mod. Phys. \textbf{79}, 135
(2007).

\bibitem{Fonseca} E. J. S. Fonseca, C. H. Monken, and S. P\'{a}dua, Phys.
Rev. Lett. \textbf{82}, 2868 (1999).

\bibitem{D'Angelo} M. D'Angelo, M. V. Chekhova, and Y. Shih, Phys. Rev.
Lett. \textbf{87}, 013602 (2001).

\bibitem{Walther} P. Walther, J. -W, Pan, M. Aspelmeyer, R. Ursin, S.
Gasparoni, and A. Zeilinger, Nature (London) \textbf{429}, 158 (2004).

\bibitem{mitchell} M. W. Mitchell, J. S. Lundeen, and A. M. Steinberg,
Nature(London) \textbf{429}, 161 (2004).

\bibitem{Moreva} E. V. Moreva, G. A. Maslennikov, S. S. Straupe, and S. P.
Kulik, Phys. Rev. Lett. \textbf{97}, 023602 (2006).

\bibitem{Barreiro} J. T. Barreiro, N. K. Langford, N. A. Peters, and P. G.
Kwiat, Phys. Rev. Lett. \textbf{95}, 260501 (2005).

\bibitem{Cinelli} C. Cinelli, M. Barbieri, R. Perris, P. Mataloni, and F. De
Martini, Phys. Rev. Lett. \textbf{95}, 240405 (2005).

\bibitem{Barbieri} M. Barbieri, C. Cinelli, P. Mataloni, and F. De Martini,
Phys. Rev. A \textbf{72}, 052110 (2005).

\bibitem{Adamson} R. B. A. Adamson, L. K. Shalm, M. W. Mitchell, and A. M.
Steinberg, Phys. Rev. Lett. \textbf{98}, 043601 (2007).

\bibitem{Tsujino} K. Tsujino, H. F. Hofmann, S. Takeuchi, and K. Sasaki,
Phys. Rev. Lett. \textbf{92}, 153602 (2004).

\bibitem{Eisenberg} H. S. Eisenberg, J. F. Hodelin, G. Khoury, and D.
Bouwmeester, Phys. Rev. Lett. \textbf{96}, 160404 (2006).

\bibitem{Ou99} Z. Y. Ou, J. -K. Rhee, and L. J. Wang, Phys. Rev. A \textbf{60%
}, 593 (1999).

\bibitem{Ou05} Z. Y. Ou, Phys. Rev. A \textbf{72}, 053814 (2005).

\bibitem{Sun07} F. W. Sun, B. H. Liu, Y. F. Huang, Y. S. Zhang, Z. Y. Ou,
and G. C. Guo, Phys. Rev. A \textbf{76}, 063805 (2007).

\bibitem{Greenberger} D. M. Greenberger, M. A. Horne, and A. Zeilinger, in
\textit{Bell's Theorem, Quantum Theory, and Conceptions of the Universe},
edited by M. Kafatos (Kluwer Academics, Dordrecht, The Netherlands, 1989),
pp. 73--76.

\bibitem{Bollinger} J. J. Bollinger, W. M. Itano, D. J. Wineland, and D. J.
Heinzen, Phys. Rev. A \textbf{54}, R4649 (1996).

\bibitem{Sun06} F. W. Sun, Z. Y. Ou, and G. C. Guo, Phys. Rev. A \textbf{73}%
, 023808 (2006).

\bibitem{Sun06PRA} F. W. Sun, B. H. Liu, Y. F. Huang, Z. Y. Ou, and G. C.
Guo, Phys. Rev. A \textbf{74}, 033812 (2006).

\bibitem{note} Actually, different DOF can be combined into one DOF in
mathematics just to span the basis.

\bibitem{Loudon} R. Loudon, \textit{The Quantum Theory of Light} (Oxford
University Press, New York, 2000).

\bibitem{Resch} K. J. Resch, \textit{et al.}, Phys. Rev. Lett. \textbf{98},
223601 (2007).

\bibitem{HBT} R. Hanbury Brown and R. Q. Twiss, Nature(London) \textbf{177},
27 (1956).

\bibitem{Sun07PRA} F. W. Sun, \textit{et al.}, Phys. Rev. A \textbf{76},
052303 (2007).
\end{thebibliography}
\end{document}